\documentstyle[twoside,fleqn,espcrc2]{article}
\newcommand{\lef}{(1-\gamma_5)}
\newcommand{\rig}{(1+\gamma_5)}

\newcommand{\ded}{(\delta^d_{12})}

\newcommand{\J}{\hat{J}}
\newcommand{\U}{\hat{U}}

\newcommand{\W}{\hat{W}}

\renewcommand{\d}{\delta}

\newcommand{\eps}{\epsilon}
\newcommand{\DFz}{\Delta F=0}

\newcommand{\as}{\alpha_{ s}}
\newcommand{\MSbar}{\overline{\mbox{MS}}}
\newcommand{\gammaz}{\hat{\gamma}^{(0)T}}
\newcommand{\gammau}{\hat{\gamma}^{(1)T}}
\newcommand{\bea}{\begin{eqnarray}}
\newcommand{\eea}{\end{eqnarray}}
\newcommand{\be}{\begin{equation}}
\newcommand{\ee}{\end{equation}}
\newcommand{\nn}{\nonumber}
\newcommand{\fr}{\frac}

\title{NLO corrections to $\Delta F=2$ effective Hamiltonians and
SUSY contribution to FCNC}
\author{Ignazio Scimemi\address{Dip. di Fisica,
Universit\`a degli Studi di Roma ``La Sapienza",
P.le A. Moro 2, 00185 Roma, Italy.}}
\begin{document}
\begin{abstract}
The most general QCD-NLO Anomalous-Dimension Matrix  of all four-fermion
dimension six $\Delta F=2$ operators is presented.
Two applications of this  Anomalous-Dimension Matrix
to the study of SUSY contribution
to $K-\bar K$ mixing are also discussed. 
\end{abstract}
\maketitle
%\twocolumn
\section{Introduction}
\label{sec:intro}
Theoretical predictions of several measurable quantities, which are relevant 
in $K$-, $D$- and $B$-meson phenomenology, depend  on the matrix 
elements of some $\Delta F=2$ four-fermion operators. 
Examples are given by FCNC effects in SUSY extensions of the Standard 
Model~\cite{pellicani}-\cite{beautysusy2}, 
by the $B^0_s$--$\bar B^0_s$ width difference~\cite{bbd}
and by the $O(1/m_b^3)$ corrections in inclusive b-hadron decay rates
 (which actually depend on the matrix elements
of several four-fermion $\DFz$ operators~\cite{ns}). 
In all these cases, the relevant
operators have the form 
\be
 Q= C^{\alpha\beta\rho\sigma} ({\bar
b}_{\alpha}\Gamma q_{\beta})
  ({\bar b}_{\rho} \Gamma q_{\sigma}) \, , 
  \label{eq:general} 
\ee
where $\Gamma$ is a generic Dirac matrix acting on (implicit) spinor
indices; $\alpha$--$\sigma$ are colour indices and 
$C^{\alpha\beta\rho\sigma}$ is either $\delta^{\alpha\beta}\delta
^{\rho\sigma}$ or $\delta^{\alpha\sigma}\delta
^{\rho\beta}$. 
All the  operators discussed in this paper  appear 
in some ``effective" theory, obtained by using the Operator
Product Expansion (OPE).
As a consequence, in all cases, three steps are necessary for obtaining 
physical amplitudes from their matrix elements:

i) matching of the original theory to the effective one
at some large energy scale; 

ii) renormalization-group evolution from the large energy scale 
to a low scale suitable for the calculation of the hadronic matrix elements 
(typically $1$--$5$~GeV);

iii) non-per\-tur\-ba\-ti\-ve cal\-cu\-la\-tion of the had\-ron\-ic 
mat\-rix elements.

In the following  a calculation of the two-loop
Anomalous Dimension Matrix (ADM) relevant for $\Delta F=2$ 
transition amplitudes  and its application to FCNC in SUSY extension of the 
Standard Model (SM) is presented.
I refer to the original papers for all details of the calculations, {\it i.e.},
respectively to ref.~\cite{pap1} for the calculation of the ADM and to 
ref.~\cite{pap2,pap3} for its applications to  FCNC effects in Supersymmetry.

The paper is organized as follows. In sec.~\ref{sec:heff} I introduce 
the  operator basis, the Wilson coefficients and the Renormalization Group
Equations (RGE). In sec.~\ref{sec:deltas} constraints on 
SUSY parameters coming from FCNC  in the $K-\bar K$ mixing are derived.
\section{Effective Hamiltonian  formalism and ADM}
\label{sec:heff}

  The three steps (i-ii-iii, sec.\ref{sec:intro}) needed to use  OPE
   are treated  in the following. 

In all cases of interest, the matrix elements of the effective Hamiltonian 
can be written as
\be 
\langle F \vert {\cal H}_{eff}\vert I \rangle = \sum_{i}
\langle F \vert Q_{ i}(\mu) \vert I \rangle C_{ i}(\mu) 
\, , \label{wope} 
\ee
where the $Q_i(\mu)$s are the relevant operators renormalized at the 
scale $\mu$ and the $C_i(\mu)$s are the corresponding Wilson coefficients.

The  set of operators  relevant
for a  description  of FCNC effects in SUSY models is:
\bea 
Q_1 &=& \bar d^\alpha \gamma_\mu \lef s^\alpha \ \bar d^\beta \gamma_\mu \lef 
s^\beta\, , \nn \\ 
Q_2 &=& \bar d^\alpha \lef s^\alpha \ \bar d^\beta \lef s^\beta \, , \nn \\ \nn
Q_3&=& \bar d^\alpha \lef s^\beta \ \bar d^\beta \lef  s^\alpha \, , \\ 
Q_4 &=& \bar d^\alpha \lef s^\alpha \ \bar d^\beta \rig s^\beta \, , \nn \\ 
Q_5&=& \bar d^\alpha \lef s^\beta \ \bar d^\beta \rig s^\alpha\, ,  
\label{eq:susy} 
\eea 
together with  operators $\tilde Q_{1,2,3}$ which
 can be obtained from $Q_{1,2,3}$ by
the exchange $\lef \leftrightarrow \rig$.

Let us now represent  the operators as row vectors $\vec Q$
 and the coefficients, $\vec C(\mu)$, as 
column ones. The vectors $\vec C(\mu)$ are
expressed in terms of their counter-part,
computed at a large scale $M$, through 
the renormalization-group evolution matrix $\W[\mu,M]$
\be \vec C(\mu) = \W[\mu,M] \vec C(M)\, . \label{evo} \ee
The initial conditions for the evolution equations,
$\vec C(M)$, are obtained by matching the full theory, which includes
propagating heavy-vector bosons ($W$ and $Z^0$), 
the top quark, SUSY particles, etc.,
to the effective theory where the $W$,
$Z^0$, the top quark and all the heavy particles
have been removed. In general,
$\vec C(M)$ depend on the definition of the operators
in a given renormalization scheme. 

The coefficients $\vec C(\mu)$ obey the ren\-or\-mal\-iz\-ation-group 
equa\-tions:
\bea 
\Bigl[ - \frac {\partial} {\partial t} + \beta ( \as )
\frac {\partial} {\partial \as} + \beta_\lambda( \as )
\lambda\frac {\partial} {\partial \lambda} -\Bigr. 
& & \nn \\ 
\Bigl.
\frac {\hat \gamma^T ( \as) }{2} \Bigr] 
\vec C(t, \as(t), \lambda(t)) &=&0 \, , 
\label{rge} 
\eea
where $t=\ln ( M^2 / \mu^2 )$.  The term proportional to $\beta_\lambda$,
the $\beta$-function of the gauge parameter $\lambda(t)$ (for covariant gauges),
takes into account the gauge dependence of the Wilson coefficients 
in gauge-dependent renormalization schemes, such as the 
RI scheme~\cite{Ciuchini2}.
%~\footnote{ In the following, I will denote by $\lambda=1$ the Feynman gauge
%and $\lambda=0$ the Landau gauge.}
 This term  is absent in standard 
 $\MSbar$ schemes, independently of the regularization 
which is adopted (NDR, HV or DRED for example). 

If no quark threshold between the scales $M$ and $\mu$ is crossed,
eq.~(\ref{rge}) can be easily solved (the formulae for the general case
can be found in refs.~\cite{Ciuchini,bjl}).
At the next-to-leading order, one can write
\bea
\W[\mu,M]  = \hat M[\mu] \U[\mu, M] \hat M^{-1}[M] \, , 
 \label{monster} \eea
where $\U$ is the leading-order evolution matrix
\be
\U[\mu,M]=  \left[\frac{\as (M)}{\as (\mu)}\right]^{
      \gammaz / 2\beta_{ 0}} \, ,
\label{u0} \ee
and the NLO matrix is given by
\be \hat M[\mu] =
 \hat 1 +\frac{\as (\mu)}{4\pi}\J[\lambda(\mu)] \, .
\label{mo2} \ee

By substituting the expression of the $\vec C(\mu)$  given in 
eq.~(\ref{evo}) in the re\-nor\-ma\-li\-za\-tion-group equations (\ref{rge}), 
and using $\hat W[\mu, M]$ written as in eqs.~(\ref{monster})--(\ref{mo2}),
one finds that the matrix $\J$ satisfies the equation
\be
\J+\frac{\beta^0_\lambda}{\beta_0}
\lambda \frac{\partial \J}{\partial \lambda}
- \left[\J,\frac{\gammaz}{2\beta_{ 0}}\right] =
     \frac{\beta_{ 1}}{2\beta^2_{ 0}}\gammaz-
     \frac{\gammau}{2\beta_{ 0}} \, . \label{jj}
\ee
In eqs. (\ref{u0}) and (\ref{jj}), $\beta_{ 0}$,  $\beta_{ 1}$
and $\beta^0_\lambda$ are the first coefficients of the 
$\beta$-functions of $\as$ and of $\lambda$, respectively; 
$\hat \gamma^{(0)}$ and $\hat \gamma^{(1)}$ are the LO and NLO anomalous 
dimension matrices. 
$\hat U$ is determined by the LO anomalous
dimension matrix $\hat \gamma^{(0)}$ and 
is therefore regularization and renormalization-scheme independent; 
 the two-loop anomalous dimension matrix $\hat \gamma^{(1)}$, 
and consequently $\J$ and $\hat W[\mu , M]$, are, instead, 
renormalization-scheme dependent.
The precise definition and 
a complete discussion of the scheme dependence of 
eq.~(\ref{jj}) can be found in ref.~\cite{pap1}
and in references therein.

Eventually, the B-parameters which par\-am\-et\-rize the had\-ron\-ic matrix 
elements of eq.~(\ref{eq:susy})
have been evaluated non-perturbatevely on the lattice~\cite{allt,gupta}.

%Eventually for what concerns
%hadronic matrix elements  of eq.~(\ref{wope}),
%they can be evaluated non-perturbatively introducing
% the so-called B-pa\-ra\-me\-ters.
%I refer to refs.~\cite{pap2,pap3,allt} for  their  precise definition
%for the operator basis of eq.~(\ref{eq:susy}) and to refs.~\cite{allt,gupta}
%for their extimation on lattice.
In what follows I have used the extimate of ref.~\cite{allt}.
 
 \section{FCNC in SUSY}

\label{sec:deltas}
The considered operator basis is the one relevant for the analysis of FCNC   
in SUSY  in the so-called mass insertion approximation~\cite{hall}.
In this approach for  fermions and sfermions states
 all the couplings of these particles to neutral gauginos are
  flavor diagonal and FCNC
effects are shown by the non-diagonality of  sfermion propagators.
The pattern of flavor change, for the K-system, is given by the ratio 
\be
(\delta^{d}_{ij})_{AB}= 
\fr{(m^{\tilde d }_{ij})^2_{AB}}{M_{sq}^2} \ ,
\ee
where $ (m^{\tilde d }_{ij})^2_{AB}$  are the off-\-diag\-on\-al 
el\-em\-ents  of 
the
$\tilde d $ mass squared mat\-rix that mix\-es
 fla\-vor $i$, $j$ for both left- and
right-han\-ded scal\-ars ($A,B=$Left, Right), and 
$M_{sq}$ is an av\-er\-age squark mass, see {\it e.g.}~\cite{pellicani}.
The sfer\-mion propagators are expanded as a series in terms of 
the $\delta$'s and the
contribution of the first term of this expansion is considered.
Assuming that all s-particles are heavier than fermions one produces
  effective four fermions interactions and OPE can be used.

The Wilson coefficients at the matching scale $M_{sq}$ can be found, {\it e.g.},
in ref.~\cite{pellicani,pap2,pap3,bagg} and I refer to these articles for
their complete expression.

One can provide a  
set of constraints on SUSY variables coming from the $K_L-K_S$
mass difference, $\Delta M_K$ and the CP violating 
parameter $\eps_K$ defined as
\bea
\Delta M_K &=&2 {\rm{Re}}\langle K^0|H_{\rm{eff}}|\bar{K}^0\rangle, \nn \\
\eps_K &=& \fr{1}{\sqrt{2} \Delta M_K}{\rm{Im}}
\langle K^0|H_{\rm{eff}}|\bar{K}^0\rangle.
\eea
The parameter space  is  composed of two real and four complex entries, that
is $M_{sq}$, $m_{\tilde g}$ and $\ded_{LL}$, $\ded_{LR}$, $\ded_{RL}$, 
$\ded_{RR}$.

Neglecting interference
among different SUSY contribution one  gives upper bound on the $\d$'s,  
giving specific values
to $M_{sq}$ and $m_{\tilde g}$.
I consider the conditions  $M_{sq} \sim m_{\tilde g}$
in subsec.~\ref{subsec:pap2}  and the case
in which the scalars of the first-two
 generations are heavier than the rest
of the supersymmetric spectrum  in
 subsec.~\ref{subsec:pap3}.
   Constraints on individual $\d$'s are provided. 
Indeed,  it is meaningful to study the interference
of  cancellation effects  only in specific models.

The physical condition used to get
the bounds on the $\delta$' s is  that the SUSY contribution
 (proportional to  each single $\d$)
 plus the SM
contribution to $\Delta M_K$ and $\eps_K$ do not exceed the experimental value
of these quantities.

\subsection{$M_{sq} \sim m_{\tilde g}$}
\label{subsec:pap2}

For Left-Right mass insertions, I consider two possible (extreme)
cases: 
$|\ded_{LR}|\gg|\ded_{RL}|$ and $|\ded_{LR}|=|\ded_{RL}|$.
In the second case the contributions proportional to 
$\ded_{LR}^2$, $\ded_{RL}^2$ and $\ded_{LR}\ded_{LR}$
are combined.
This approach improves the one of refs.~\cite{bagg,hage}, 
where contributions proportional to 
 $\ded_{LR}\ded_{LR}$
were considered independently from the ones proportional to
$\ded_{LR}^2$ and $\ded_{RL}^2$.

$\ded_{LL}\ded_{RR}$ contribution  can be treated independently from
the $\ded_{LL}^2$ and $\ded_{RR}^2$ ones, since
the latter ones do not generate Left-Right operators and they are 
 suppressed with respect to the first contribution.
 
Results for the bounds of the $\d$'s for $M_{sq}\sim 200$ GeV and various values
of the ratio $x= m_{\tilde g}^2/M_{sq}^2$ derived for real parts of the
$\d$'s are shown in tab.~\ref{tab:m200}.
Let us compare these results with the  previous analyses of 
refs.~\cite{pellicani,bagg}.
In both papers the matrix elements were computed in 
the Vacuum Insertion Approximation (VIA) and with no evolution~\cite{pellicani}
 or leading  order evolution~\cite{bagg} of the coefficient functions.
 In this analysis lattice B-parameters and NLO evolution of the coefficients
 have been calculated.
 This allows  a consistent matching of the renormalization-scheme
 and $\mu$ dependence of the QCD-corrected ${\cal{H}}^{\Delta S=2}_{eff}$.
 
 \subsection{Heavy first-two generation scalars}
\label{subsec:pap3}

Let us assume now that the scalars of the first two generations
(the typical mass of which is here again denoted by $M_{sq}$)
is much heavier than the rest of the SUSY spectrum
(which has a typical mass of $m_{\tilde g}$).
In this case  the  heavy squarks   have to be  
 integrated out as a first, at a scale $M_{sq}$. 
 Then, the coefficients must 
be evolved to the scale $m_{\tilde g}$ where
 all the rest of the supersymmetric
particles are also integrated out.

In table~\ref{tab:m250} results are shown for the real parts
of the $\delta$'s with   $m_{\tilde g}\sim 250$ GeV and
several values of $M_{sq}$.

The constraints that come from the four possible insertions 
of the $\d$'s are presented:
in the first and second rows only
 terms proportional respectively to $\ded_{LL}$ and
$\ded_{LR}$ are considered;
 in the last two rows the contribution of  operators with
opposite chirality, $RR$ and $RL$, is also evaluated by assuming
$\ded_{LL}=\ded_{RR}$ and $  \ded_{LR}=\ded_{RL}$.

The theoretical improvements of this results with respect to ref.\cite{bagg} have already
been discussed in subsec.~\ref{subsec:pap2}.
The combination of B-parameters and NLO-QCD corrections change 
the LO-VIA results by 
  about $25-35\%$. As expected~\cite{bagg}, the tightest constraints are for
the cases $\ded_{LL}=\ded_{RR}$ and $\ded_{LR}=\ded_{RL}$.
 In these cases the coefficients proportional to $\ded_{LL}\ded_{RR}$,
$\ded_{LR}\ded_{RL}$ dominate the others.
\section{Acknowledgements}
I wish to thank all  the authors of refs.~\cite{pap1,pap2,pap3}
with whom the results presented in this work have been obtained.

Particular thanks to Stephan Narison for the beautiful conference
he has organized.
{\small
 \begin{table} 
 \begin{tabular}{|c|c|c|c|c|}  \hline  
 & No-QCD,  & LO,  & LO,  & NLO,   \\
 &VIA       &VIA   &$B_i$ & $B_i$  \\
 &$\times 10^{-3}$&$\times 10^{-3}$&$\times 10^{-3}$&$\times 10^{-3}$\\
 \hline \hline
 $x$ & \multicolumn{4}{c|}{$\sqrt{|\rm{Re}  (\delta^{d}_{12})_{LL}^{2}|} $} \\ 
\hline 
 0.3& $5.0$ 

& $5.7$ 

& $7.7$ 

& $7.7$ 
\\ 
1.0& $11$ 

& $12$ 

& $16$ 

& $16$ 
\\ 
4.0& $25$ 

& $29$ 

& $39$ 

& $39$ 
\\ 
\hline  \hline
 $x$ & \multicolumn{4}{c|}{$\sqrt{|\rm{Re}  (\delta^{d}_{12} )_{LR}^{2}|} 
 \qquad (|(\delta^{d}_{12})_{LR}|\gg 
|(\delta^{d}_{12})_{RL}|)$} \\ 
\hline 
 0.3& $1.1$ 
 
& $0.84$ 
 
& $1.1$ 
 
& $0.96$ 
\\ 
1.0& $1.2$ 
 
& $0.93$ 
 
& $1.2$ 
 
& $1.1$ 
\\ 
4.0& $1.8$ 
 
& $1.3$ 
 
& $1.6$ 
 
& $1.5$ 
\\ 
\hline  \hline
 $x$ & \multicolumn{4}{c|}{$\sqrt{|\rm{Re}  (\delta^{d}_{12} )_{LR}^{2}|} 
\qquad ((\delta^{d}_{12})_{LR} = 
 (\delta^{d}_{12} )_{RL})$} \\ 
 \hline 
 0.3& $2.0$ 
 
& $1.4$ 
 
& $0.89$ 
 
& $0.67$ 
\\ 
1.0& $1.1$ 
 
& $0.97$ 
 
& $1.8$ 
 
& $3.0$ 
\\ 
4.0& $1.3$ 
 
& $1.0$ 
 
& $1.4$ 
 
& $1.3$ 
\\ 
\hline \hline 
 $x$ & \multicolumn{4}{c|}{$\sqrt{|\rm{Re}  (\delta^{d}_{12} )_{LL}
(\delta^{d}_{12})_{RR}|} $} \\ 
 \hline 
0.3& $0.64$ 
 
& $0.39$ 
 
& $0.40$ 
 
& $0.33$ 
\\ 
1.0& $0.71$ 
 
& $0.44$ 
 
& $0.45$ 
 
& $0.37$ 
\\ 
4.0& $1.0$ 
 
& $0.61$ 
 
& $0.62$ 
 
& $0.52$ 
\\ 
\hline  \hline 
 \end{tabular} 
 \caption{Limits on $\mbox{Re}\left(\delta_{ij} 
 \right)_{AB}\left(\delta_{ij}\right)_{CD}$, with $A,B,C,D=(L,R)$, for an
  average squark mass $M_{sq}=200$ GeV and for different values
 of $x=m_{\tilde{g}}^2/M_{sq}^2$.} 
 \label{tab:m200} 
\end{table} 
}
%________________________________________________________________________
{\small
\begin{table} 
\begin{center} 
\begin{tabular}{|c|c|c|c|c|} 
\hline 
 $M_{sq}$ & No-QCD,  & LO,  & LO,      & NLO,  \\
  $[{\rm TeV}]$  &VIA       & VIA & $B_i$ & $B_i$  \\
    &$\times 10^{-2}$ &$\times 10^{-2}$ &$\times 10^{-2}$ &$\times 10^{-2}$ \\
 \hline \hline
 \multicolumn{5}{|c|}{$\sqrt{|{\rm{Re}} (\delta^d_{12})^2_{LL}|}$} 
 \\ 
 \hline 
 2 & $3.1$ & $3.6$ & $4.9$ & $4.9$ \\ 
 5 & $7.5$ & $8.8$ & 0.12 & 0.12\\ 
 10 & 15 & 18&25 & 24\\ 
\hline \hline 
\multicolumn{5}{|c|}{$\sqrt{|{\rm{Re}} (\delta^d_{12})^2_{LR}|}$} \\ 
\hline 
2 & $2.1$ & $1.4$ & $1.8$ & $1.6$ \\ 
 5 & $9.8$ & $6.5$ & $8.2$ & $7.2$\\ 
 10 & 34 & 22&28 & 25\\ 
\hline \hline 
\multicolumn{5}{|c|}{$\sqrt{|{\rm{Re}} 
(\delta^d_{12})_{LR}={\rm{Re}} (\delta^d_{12})_{RL}|}$} \\ 
\hline 2 & $0.66$ & $0.35$ & $0.35$ & $0.28$ \\ 
 5 & $1.5$ & $0.77$ & $0.79$ & $0.64$\\ 
 10 & $3.0$ & $1.5$&$1.5$ & $1.2$\\ 
\hline \hline 
\multicolumn{5}{|c|}{$\sqrt{|{\rm{Re}} (\delta^d_{12})^2_{LL}={\rm{Re}} 
(\delta^d_{12})^2_{RR}|}$} \\ 
\hline 
2 & $1.1$ & $0.52$ & $0.51$ & $0.41$ \\ 
 5 & $4.1$ & $1.6$ & $1.6$ & $1.3$\\ 
 10 & 10 & $3.6$&$3.4$ & $2.7$\\ 
\hline \hline 
\end{tabular} 
\caption{{ Limits on ${\rm Re }  (\delta^d_{12})_{AB}$ from 
$\Delta M_K$ with gaugino masses of 250   GeV.}}
\label{tab:m250}
\end{center} 
\end{table} }


\begin{thebibliography}{88}

\bibitem{pellicani} E. Gabrielli, A. Masiero and L. Silvestrini,
Phys.~Lett. B374 (1996) 80; 
F. Gabbiani, E. Gabrielli, A. Masiero and L. Silvestrini,
Nuc.~Phys. B477 (1996) 321 and refs. therein.

\bibitem{beautysusy2} J.A. Bagger, K.T. Matchev and R.J. Zhang,
preprint JHU-TIPAC-97011, hep-ph/9707225 and refs. therein.

%\bibitem{lr} G. Beall, M. Bander and A. Soni,
%Phys.~Rev.~Lett. 48 (1982) 848.

\bibitem{bbd} M.~Beneke, G.~Buchalla and I.~Dunietz,
Phys.~Rev. D54 (1996) 4419 and refs. therein.

\bibitem{ns} M.~Neubert and C.T.~Sachrajda, 
Nucl.~Phys. B483 (1997) 339 and refs. therein. 

\bibitem{pap1} M. Ciuchini et al. Nuc.~Phys. B523 (1998) 501.
\bibitem{pap2} M. Ciuchini et al. hep-ph/9808328.
\bibitem{pap3} R. Contino and I. Scimemi, preprint ROME1-1216/98.  
\bibitem{hall} L.~J.~Hall, V.~A.~Kostolecki and S.~Raby, Nucl. Phys. B267 (1986) 415.
\bibitem{bagg} J.~A.~Bagger, K.~T.~Matchev and R.~J.~Zhang,
               Phys. Lett. B412 (1997) 77.
\bibitem{Ciuchini2} M.~Ciuchini {\it et al.}, Z.~Phys. C68 (1995) 239.
\bibitem{Ciuchini} M.~Ciuchini, E.~Franco, G.~Martinelli and L.~Reina,
                  Nucl.~Phys. B415 (1994) 403.
\bibitem{bjl} A.J.~Buras, M.~Jamin, M.E.~Lautenbacher and P.H.~Weisz,
Nucl.~Phys. B370 (1992) 69; Addendum, Nucl.~Phys. B375 (1992) 501.
\bibitem{allt} C. R. Allton et al. hep-lat/9806016
\bibitem{gupta} R. Gupta, T. Bhattacharya and S. Sharpe, Phys. Rev. D55 (1997)
                  4036.
\bibitem{hage} J.S. Hagelin, S. Kelley and T. Tanaka, 
              Nucl. Phys. B322 (1989) 55.
\end{thebibliography}
\end{document}